\documentclass[aps,amsfonts,nofootinbib,twocolumn]{revtex4-1}
\usepackage{amsmath}
\usepackage{amsfonts}
\usepackage{amssymb}

\def\openone{\leavevmode\hbox{\small1\kern-3.8pt\normalsize1}}
\def\N{\leavevmode\hbox{ Z \kern-8 pt\normalsize{Z}}}
\def\openone{\leavevmode\hbox{\small1\kern-3.8pt\normalsize1}}
\def\openJ{\leavevmode\hbox{J \kern-9.5pt\normalsize J}}
\def\openS{\leavevmode\hbox{ S \kern-9.3pt\normalsize S}}
\newcommand{\bb}{\begin{equation}}
\newcommand{\ee}{\end{equation}}
\newcommand{\eqb}{\begin{eqnarray}}
\newcommand{\eqf}{\end{eqnarray}}
\newcommand{\nc}{{\mbox{\tiny{NC}}}}

\newcommand{\Landau}{\mbox{\tiny{Landau}}}
\usepackage{color}
\begin{document}

\title{Dynamics determines Geometry}
\author{Sergio A. Hojman}
\email{sergio.hojman@uai.cl}
 \affiliation{Departamento de Ciencias,
Facultad de Artes Liberales, Facultad de Ingenier\'{\i}a y Ciencias,
Universidad Adolfo Ib\'a\~nez, Santiago, Chile,\\ and Departamento
de F\'{\i}sica, Facultad de Ciencias, Universidad de Chile,
Santiago, Chile,\\ and Centro de Recursos Educativos Avanzados,
CREA, Santiago, Chile.}

\author{J.     Gamboa     }
\email{jgamboa55@gmail.com} 
\affiliation{Departamento   de  F\'{\i}sica,
  Universidad  de  Santiago de  Chile,  Casilla  307, Santiago,  Chile
  \\ and \\ Facultad de F\'{\i}sica, Pontificia Universidad Cat\'olica
  de Chile, Santiago, Chile} 

\author{F.     M\'endez}
\email{fernando.mendez.f@gmail.com} 
\affiliation{Departamento      de
  F\'{\i}sica,  Universidad   de  Santiago  de   Chile,  Casilla  307,
  Santiago, Chile}

\begin{abstract}
The inverse problem of  calculus of variations and $s$-equivalence
are re-examined  by using  results obtained from non-commutative
geometry ideas. The  role played by the structure of the modified
Poisson brackets is discussed in a general  context and it is argued
that classical $s$-equivalent systems may be non-equivalent at the
quantum mechanical level. This last fact is explicitly discussed
comparing different approaches to deal with the Nair-Polychronakos
oscillator.
\end{abstract}


\maketitle

\section{Introduction}

Consider a system $S$ of  $n$ differential equations for $n$ variables
$q^i$. The Inverse Problem of  the Calculus of Variations (IPCV) deals
with  the   question  of  existence  and   uniqueness  of  variational
principles from which the system $S$ may be derived.

If at  least one  Lagrangian for  system $S$ exists  then the  task of
constructing   one   (or    several)   variational   principle(s)   or
Lagrangian(s) is also  part of the IPCV. This  means that variation of
the action constructed from  one of those Lagrangian functions, yields
the original  system $S$  or a  system $S'$ equivalent  to it,  in the
sense that  the space of solutions  of $S$ and $S'$  are identical. If
this  is  the  case,  systems  $S$  and  $S'$  are  called  ``solution
equivalent'' or ``$s$-equivalent''.

As far  as we are aware,  the first significant  contributions to this
field were  made by  Helmholtz in 1887  \cite{helm} and by  Darboux in
1894 \cite{darb} for second order differential equations.  Helmholtz
found  the conditions  for the  existence of  a Lagrangian  written in
terms of the (second order) differential equations of the system $S$. If
Helmholtz  conditions are  satisfied,  then a  Lagrangian which  (upon
variation of  its action integral)  yields (exactly) the  second order
differential equations of system $S$ exists.

A  little bit later,  Darboux, solved  completely the  one dimensional
($n=1$) case  showing, in so doing,  that in one  dimension a Lagrangian
(for one second  order differential equation) always exists  and it is
not unique in a non trivial fashion.

There  is,  of  course,  the  familiar non  uniqueness  of  Lagrangian
functions which stems from the  addition of a total time derivative of
an  arbitrary  function.  In  the  one  dimensional  case,  there  are
Lagrangian functions  which give rise to (infinitely  many) systems of
second order  differential equations  which are s-equivalent  (but not
identical) to system $S$.

These  variational principles give  rise to  different sets  of second
order  differential equations which  have the  same set  of solutions.
The   (first  order   differential  equations)   Hamiltonian  theories
constructed from these Lagrangian formulations are different from each
other  in the  sense  that  they give  rise  to different  Hamiltonian
functions and different Poisson Brackets relations.

In  1941,   Douglas  \cite{doug}  solved  the   two  dimensional  case
completely. Three possible outcomes arise in the two dimensional case:
a) no Lagrangian exists, b) there  is exactly one (up to addition of a
total  time derivative)  Lagrangian or  c) there  are  infinitely many
Lagrangian functions  for the system of two  second order differential
equations.

In the case of  first order differential equations, Havas
\cite{havas} made progress towards the solution of  this problem
which was completely solved  by Hojman  and  Urrutia in  1981
\cite{hu},  who  provided  a   way  of  constructing  infinitely
many Lagrangians for  such systems  and presented examples of first
order Lagrangians for systems for which no second order Lagrangians
exist (in this context see also \cite{sh}, \cite{hs}).

It is then clear that the quantization of such systems might give
rise to different  quantum theories\footnote{In this paper  we will
discuss the first quantization  scheme only.} --  not only  due to
the well known problems  of operator  ordering -- since different
Hamiltonian (and Poisson Brackets) structures give rise to the same
classical equations of motion. These Hamiltonian structures cannot
be related by canonical transformations.

One could argue that at the end, among  all these Hamiltonian
structures which, in principle, might give  rise to different
quantum theories, only those whose predictions are realized in
nature -- and verified through experiments -- are of physical
interest. This is true, but here, an implicit assumption is made: nature
selects only one (out of infinitely many, in some cases)
Hamiltonian structure and discards the rest of them.

A counterexample of this  can be found in noncommutative spaces
which have attracted much attention recently.  Its connection with
different problems in physics  have been widely discussed in  the
literature, in the   context   of   string   theory   \cite{string},
field   theory \cite{fields} and gravitation \cite{gravitation}.

Noncommutative quantum  mechanics (NCQM), on the other  hand, has
also been explored as a possible scenario to test the physical
consequences of  this new  structure proposed  for space
\cite{ncqm}.  The two dimensional  case,  for  instance,  can  be
solved  for  any  central potential \cite{we1}, and it is  possible
to show that this problem is connected with the Landau problem for
the lowest energy levels.

From a different point of view, NCQM has proven to be good
laboratory to test new approaches which might shed light on long
standing problems in physics. In  \cite{we2}, for instance, a new
kind of noncommutativity  which incorporates the spin of the
particles has been proposed as an alternative mechanism to explain
superconductivity involving triplet states. In a recent work
\cite{graphene}, its relation with the physics of graphene is
explored.

In the present paper we  consider a particle whose classical
equations of motion  (EoM) are a  generalization of the  EoM of a
particle  in a magnetic field.  We construct two Hamiltonian
structures  for these classical EoM which give rise to two different
and inequivalent quantum theories. One of them is related to the
Nair-Polychronakos anisotropic harmonic oscillator \cite{np}, while
the  other one is described by a Hamiltonian which is the addition
of a harmonic oscillator Hamiltonian plus a term proportional to
angular momentum.\\

A classical set of second order equations of motion $S$ is said to
be non Lagrangian if there exists no Lagrangian which yields second
order equations which are (identical or at least) equivalent to the
set $S$. Nevertheless, a Hamiltonian structure which produces first
order equations equivalent to $S$ may always be constructed using
different approaches \cite{hu}, \cite{hs}, \cite{sh2}. The
Hamiltonian structures for non Lagrangian systems are always
noncommutative in the sense that the space coordinates Poisson
Brackets sub-matrix  does not vanish \cite{sh}, \cite{hs}. We
illustrate this point with some examples.

In order to prove the previous assertions, we will briefly  review
the construction of first order Lagrangian, Poisson Brackets and
Hamiltonian structures in the last part of this section. In section
II, we explore different first order and second order Lagrangian
structures that give rise to equivalent sets of equations of motion
and their relation to noncommutative spaces. Section  III is devoted
to the quantum mechanical discussion of the model and, in the last
section, discussion and conclusions are presented.

Let us start by considering\footnote{We consider an even dimensional
phase space for simplicity, which does not mean that odd dimensional
phase spaces cannot be defined.} a set of coordinates in phase space
$\{x^a\}$, with $a\in \{1,2,\cdots,2n\}$. A first order Lagrangian
$L$ is  the most general Lagrangian such that its Euler-Lagrange
equation are first order, namely \cite{hu}
\begin{equation}
\label{fol} L=\ell_a(x^b)\dot{x}^a+\ell_0(x^b).
\end{equation}
In fact, its (first order) Euler Lagrange equations are
\begin{equation}
\label{folem}
\sigma_{ab}\dot{x}^b = \ell_{0,b},
\end{equation}
where $A,_{c}$ denotes partial derivative of $A$ with respect to
$x^c$ and
\begin{equation}
\label{sigma}
\sigma_{ab}\equiv \ell_{a,b} -\ell_{b,a}.
\end{equation}
The  exterior derivative  of the  Lagrange Brackets  two-form $\sigma$
vanishes identically, i.e.,
\begin{equation}
\label{extdersigma}
{\sigma_{ab}},_{c} + {\sigma_{bc}},_{a} + {\sigma_{ca}},_{b}\equiv 0.
\end{equation}

The Hamiltonian is straightforwardly  computed. In fact, let us define
the Poisson Brackets matrix $\openJ$ as  the inverse (up to a sign) of
the  Lagrange brackets  $\sigma_{ab}$.  The first  order equations  of
motion can be also written as
\begin{equation}
\label{foham}
\dot{x}^ a = \openJ^{ab}\frac{\partial H}{\partial x^ b},~~~~~~~~~~~~
\sigma_{ab}\openJ^{bc} = -\delta_a^{c}
\end{equation}
where $H$ is the Hamiltonian of the system ($H=-\ell_0$).

The previous equations may be rewritten as
\begin{equation}
\label{ham2}
\dot{x}^a = [x^a,H],~~~~~~\mbox{with}~~~~~~
[A,B]=\partial_a A\, \openJ^{ab}\,\partial_b B,
\end{equation}
where $[A,B]$ are the Poisson Brackets relations for any two
dynamical variables $A(x^a), B(x^b)$.

Consider a first order system defined by
\begin{equation}
\label{foeq}
\dot{x}^a = f^a(x^b).
\end{equation}

A Hamiltonian  structure for it is  defined in terms  of a Hamiltonian
$H$  and a  Poisson Brackets  matrix  $\openJ$ such  that $\openJ$  is
antisymmetric

\begin{equation}
\label{antisymm}
\openJ^{ab} = - \openJ^{ba},
\end{equation}
it satisfies Jacobi Identity
\begin{equation}
\label{JacId}
\openJ^{ab},_{d}\openJ^{dc}        +       \openJ^{bc},_{d}\openJ^{da}
+\openJ^{ca},_{d}\openJ^{db} \equiv 0, 
\end{equation}
and generates the equations of motion, in conjunction with $H$
\begin{equation}
\label{foeq1}
f^a = \openJ^{ab}\frac{\partial H}{\partial x^ b}.
\end{equation}

Given a Hamiltonian structure for a first order system then the
first order Lagrangian may be easily constructed \cite{hu}, Appendix
A of \cite{sh3}, \cite{sh4}, \cite{sh5}.

By the  same token,  given a second  order Lagrangian for  a dynamical
system, a  first order  one may be  easily constructed (the  one which
gives rise to Hamilton's equations, for instance).

Nevertheless,  the converse  is not  true, i.e.,  given a  first order
Lagrangian  it is  not always  possible  to construct  a second  order
Lagrangian  for   a  given  dynamical   system  \cite{hu},  \cite{sh},
\cite{hs}.

In the next section we will  specify the set of classical equations
of motion we are interested in and then, we will show  explicitly
two different  Lagrangian  and  Hamiltonian structures  for them and
their quantum theories.
\section{Classical equations of motion and Lagrangian structures}

Consider a system with $n$ spatial coordinates $\{ q_i \}_{i \in \{ 1,
2,\dots, n  \}}$ and the Lagrangian
\begin{equation}
\label{secord}
L_{q\dot{q}} = \frac{1}2 \left( T_{ij}\dot{q}^i\dot{q}^j +\theta_{ij}
q^i \dot{q}^j - V_{ij} q^i q^j \right),
\end{equation}
where $T_{ij},  \theta_{ij}$ and  $V_{ij}$ are constant  matrices with
the following symmetry properties
$$
T_{ij}=T_{ji},~~~~\theta_{ij}=-\theta_{ji},~~~~V_{ij}=V_{ji}.
$$

The Euler Lagrange equations are
\begin{equation}
\label{eom}
T_{ij}\,\ddot{q}^{\,j} - \theta_{ij}\,\dot{q}^j + V_{ij}\, q^j=0.
\end{equation}

Take, for  example     $n=2$,  $T_{ij}=  m\,\delta_{ij}$,
$\theta_{ij}  = \frac{e}{c} \epsilon_ {ij}B$  and $V_{ij}=0$,
equation (\ref{eom}) describes the dynamics of a  particle of mass
$m$ and electrical charge $e$ in the presence of a  constant
magnetic field $B$ orthogonal to the plane.

A first order Lagrangian may be straightforwardly constructed from
this second order Lagrangian (see, for instance Appendix A of
\cite{sh3}, \cite{sh4} and \cite{sh5} ). Define the variables $u^i$
by $u^i\equiv \dot{q}^i $, and the function $\bar L_{qu}$ (using the
$u^i$ definition into (\ref{secord})) by
\begin{equation}
\label{Fqu}
\bar L_{qu}=\frac{1}{2} \left( T_{ij} u^i\ u^j
+\theta_{ij} q^i u^j - V_{ij} q^i q^j \right).
\end{equation}

The first order Lagrangian ${\cal{L}}_{qu}$ is
\begin{equation}
\label{Lqu} {\cal{L}}_{qu}= \frac{\partial \bar L_{qu}}{\partial
u^i}(\dot{q}^i-u^i)+ \bar L_{qu},
\end{equation}
or
\begin{eqnarray}
\label{Lqu2}
{\cal{L}}_{qu}  =  \left(T_{ij}u^j  -  \frac{1}{2}  {\theta}_{ij}  q^j
\right) (\dot{q}^i-u^i)+ \nonumber 
\\
+ \frac{1}{2} \left( T_{ij} u^i\ u^j +\theta_{ij}q^i u^j - V_{ij} q^i q^j \right).
\end{eqnarray}

The canonical momenta are
\begin{equation}
\label{pu}
p_i=\frac{\partial {\cal{L}}_{qu} }{\partial \dot{q}^i}= T_{ij} u^j -
\frac{1}{2}{\theta}_{ij} q^j\,,
\end{equation}
from which
\begin{equation}
\label{up} u^i=(T^{-1})^{ij}\left( p_j +
\frac{1}{2}\theta_{jk}q^k\right).
\end{equation}

On the other hand, the Hamiltonian --  expressed also in terms of
$q^i$ and $u^j$ -- is
\begin{equation}
\label{hamuq} H_{qu}=\frac{1}{2}\left(T_{ij}u^iu^j
+V_{ij}q^iq^j\right).
\end{equation}

Now we can  rewrite the first order Lagrangian  ${\cal{L}}_{qu}$ (in a
Palatini like fashion) as 
\begin{equation}
\label{lqu} {\cal L}_{qu}=p_i\dot{q}^i-H,
\end{equation}
with (\ref{pu}) and (\ref{hamuq}) in (\ref{lqu}) we get
\begin{equation}
\label{lqufirst}
{\cal   L}_{qu}=   \left(T_{ij}u^j-\frac{1}{2}\theta_{ij}q^j   \right)
\frac{dq^i}{dt} -\frac{1}{2}\left(T_{ij} u^iu^j +V_{ij}q^iq^j\right).
\end{equation}

Varying $q$ and $u$  independently, the following first order
equations of motion are found
\begin{subequations}
\begin{eqnarray}
\label{eomqu}
\frac{d}{dt}\left(T_{ij}u^j-\frac{1}{2}\theta_{ij}q^j\right)
-\frac{1}{2} \theta_{ij}\dot{q}^j+V_{ij}q^j&=&0,
\\
T_{ij}\left(\dot{q}^j-u^j\right) &=& 0.
\end{eqnarray}
\end{subequations}
which are  equivalent to the  original equations (\ref{eom})
(provided $ det\ T_{ij} \neq 0 $) plus the definition of the
variables $u^j$.

A different set of variables may be used for the Lagrangian and
Poisson brackets relations. Let us choose $p_k$ variables defined in
(\ref{pu}). That means that
$$
\dot{q}^i= (T^{-1})^{ij}\left(p_j +\frac{1}{2}\theta_{jk}q^k\right),
$$
and therefore the Palatini like Lagrangian $p_i\dot{q}^i-H$ reads
now
\begin{eqnarray}
\label{lpq} {\cal L}_{qp}&=&p_i\dot{q}^i -\frac{1}{2} \bigg[
\left(p_i + \frac{1}{2}    \theta_{ik}q^k\right)   (T^{-1})^{ij}
\left(p_j    + \frac{1}{2} \theta_{jm}q^m \right)+ \nonumber
\\
&+& V_{ij}q^iq^j \bigg].
\end{eqnarray}

The first order  equations of motion  (varying $q$ and  $p$
independently) turn out to be
\begin{subequations}
\label{qpeom}
\begin{eqnarray}
\dot{p}_i    -    \frac{1}{2}\theta_{ij}(T^{-1})^{jk}   \left(p_k    +
\frac{1}{2} \theta_{km} q^m \right) &=& 0,
\\
\dot{q}^i   -(T^{-1})^{ij}\left(p_j  +  \frac{1}{2}   \theta_{jm}  q^m
\right) &=&0,
\end{eqnarray}
\end{subequations}
which are, of course, equivalent to (\ref{eom}).

In  summary, we  have  a second  order  Lagrangian ($L_{q\dot{q}}$
in (\ref{secord}))  which gives rise  to the  second order
equations of motion (\ref{eom}), and two  first order Lagrangians
(${\cal L}_{qu}$ in (\ref{lqufirst}) and ${\cal L}_{qp}$ in
(\ref{lpq})) which give rise to first order equations which are
equivalent to the second order ones.

The two different sets of variables  $ \{ q^i, u^j\}$ and $\{ q^i,
p_j\}$ have, of course, different Poisson Bracket relations. In
fact, compute the Lagrange bracket $\sigma_{ab}$ for
(\ref{lqufirst}), as well as its inverse (up to a sign)
$\openJ^{ab}$
\begin{equation}
\label{sigmaqu} \sigma_{ab}=\left(
\begin{array}{cc}
\theta_{ij} & T_{ij}
\\
- T_{ij} & 0
\end{array}\right),~~~
\openJ^{ab}=\left(
\begin{array}{cc}
0 & (T^{-1})^{ij}
\\
-(T^{-1})^{ij} & -(T^{-1}\,\theta\,T^{-1})^{ij}
\end{array}\right)
\end{equation}
where  $\{i,j\}\in\{1,2,\cdots,n\}$ and the coordinates of  the
phase space are $x^a  = \{ q^1, q^2, \cdots  , q^n , u^1, u^2,
\cdots , u^n \}$.

For the Lagrangian (\ref{lpq}), instead, we have the canonical
Lagrange bracket $\sigma$ and its inverse (up to a sign) $\openJ$
\begin{equation}
\label{sigmapq}
\sigma_{ab}=\left(
\begin{array}{cc}
0 & \openone
\\
- \openone & 0
\end{array}\right),~~~~~~
\openJ^{ab}=\left(
\begin{array}{cc}
0 & \openone
\\
-\openone & 0
\end{array}\right)
\end{equation}

In the next section we will exhibit three Hamiltonian systems which
give rise to the same classical equations of motion. Two of them are
constructed starting from the aforementioned first order Lagrangian
structures. The third hamiltonian structure cannot be derived from a second
order Lagrangian.

\section{Classical equations of motion and Hamiltonian structures}
A Hamiltonian  structure is  defined by (\ref{antisymm}), (\ref{JacId}) and
(\ref{foeq}).

Consider  now  the first  set  of  variables  $\{q,u\}$, used  in  the
previous  section. A Hamiltonian  system is  defined by  the following
Poisson bracket relations (\ref{sigmaqu}) and Hamiltonian function (\ref{hamuq})
\begin{eqnarray}
\label{hamqu} [q^i,q^j]=0,~~~~~~~             &&\mbox{[} u^i,u^j
\mbox{]}= -(T^{-1})^{ik}\theta_{km}(T^{-1})^{mj},\nonumber
\\
\mbox{[} q^i, u^j \mbox{]}&=&(T^{-1})^{ij},\nonumber
\\
H=\frac{1}{2}\bigg( T_{ij}u^iu^j &+& V_{ij}q^iq^j\bigg).
\end{eqnarray}
Hamilton's equations are equivalent to (\ref{qpeom}) and the second order ones
are equivalent to our starting set (\ref{eom}).

For  the  system described  in  variables  $\{q,p\}$, the  Hamiltonian
system is defined as follows
\begin{eqnarray}
\label{hamqp}
[q^i,q^j]&=&0,~~~~~~[ p_i,p_j ]=0,~~~~~~ [ q^i, p_j ]={\delta^{i}}_j,\nonumber
\\
H&=&\frac{1}{2}\bigg(\left(p_i   +    \frac{1}{2}
\theta_{ik}q^k \right)  (T^{-1})^{ij} \left(p_j   +    \frac{1}{2}
\theta_{jm}q^m \right) +\nonumber
\\
&+& V_{ij}q^iq^j\bigg).
\end{eqnarray}
for which second order equations of motion are again (\ref{eom}).

The preceding structures are two versions (using different phase
space variables) of Hamiltonian theories derived from the second
order Lagrangian (\ref{secord}).

The general results may now be applied to a special case $T_{ij} = V_{ij}$.
In this case, we can construct a third
Hamiltonian structure. In fact, let us denote the $2n$ coordinates of
phase space by $\{q^i,v_j\}$. The following Poisson brackets and
Hamiltonian function define a third Hamiltonian structure for (\ref{eom})
\begin{eqnarray}
\label{hamnc}
[ q^i, v_j ]={\delta^i}_j,~~&&~~\mbox{[} v_i,v_j \mbox{]}=0, \nonumber
\\
\mbox{[}q^i,q^j\mbox{]}&=&(T^{-1})^{ik}\theta_{km}(T^{-1})^{mj},
\\
H=\frac{1}{2}\bigg( (T^{-1})^{ij}v_i v_j &+& T_{ij} q^iq^j \bigg).
\end{eqnarray}

Hamilton's equations turn out to be
\begin{eqnarray}
\label{eomnc}
\dot{q}^i&=&(T^{-1})^{ij}\theta_{jm}q^m + (T^{-1})^{ij} v_j,
\\
\dot{v}_i&=&-T_{ij}q^j,
\end{eqnarray}
which are equivalent to (\ref{eom}) with $V = T$.

A  first  order  Lagrangian  for  these  equations  of  motion  can  be
calculated directly from our discussion in Section II, namely
\begin{eqnarray}
\label{lqv}
L_{qv}&=&v_k\dot{q}^k     +      \frac{1}{2}     \left(     T^{-1}\theta
T^{-1}\right)^{km} v_k \dot{v}_m - \\
\nonumber
&-& \frac{1}{2}\left( (T^{-1})^{ij} v_ iv_j + T_{ij} q^iq^j \right),
\end{eqnarray}
and it is a straightforward matter to prove that first order
equations of motion are
\begin{subequations}
\begin{eqnarray}
\dot{q}^i &=& (T^{-1})^{ik}v_k- \left( T^{-1}\theta T^{-1}\right)^{km}
\dot{v}_m,
\\
\dot{v}_i &=& -T_{ij}q^j.
\end{eqnarray}
\end{subequations}

It is worthwhile mentioning that this Hamiltonian structure is not
derivable from
a second order Lagrangian \cite{sh}, \cite{hs} .

In the following section we  will discuss a physical example where
these three Hamiltonian structures are considered.

\section{Landau problem and noncommutative spaces}
In  this section  we analyze two  very well  known  systems which
are special cases of the examples discussed above, namely,  the
charged particle in  an external, constant  magnetic field --
particle which, upon quantization, originates the so called Landau
Levels  -- and the  noncommutative harmonic  oscillator as  treated
by Nair  and Polychronakos  \cite{np}.  The first  one  corresponds
to  a system as the one described by variables $\{q,p\}$ or
$\{q,u\}$, while the second  one corresponds to  a system in
variables  $\{q,v\}$.

\subsection{Symmetric Gauge}

Consider, then, a  non relativistic particle with charge  $e$ and mass
$m$ in a region of constant magnetic field ${\bf B}$. In the symmetric gauge
the magnetic vector potential is
$$
{\bf A} =-\frac{1}{2} {\bf r}\times {\bf B},
$$
with  ${\bf  B}=B\hat{z}$, ($B$  constant). The  Lagrangian and the
Hamiltonian of this system are very  well known. We will only write
the Hamiltonian which is useful  for our discussion.  The
Hamiltonian turns out to be
\begin{equation}
\label{hlandau}
H=  \frac{1}{2m}   \left(p_1  +  \frac{1}{2}  e   B\,q^2  \right)^2  +
\frac{1}{2m}   \left(   p_2   -   \frac{1}{2}  eB\,   q^1\right)^2   +
\frac{^1}{2m}p_3^2,
\end{equation}
and the canonical Poisson brackets
$$
[q^i,q^j]=0,~~~~~~[q^i,p_j]=\delta^i_j,~~~~~~[p_i,p_j]=0.
$$
Since $p_3$ is a conserved quantity, the problem can be reduced to two
dimensions. This is just the Hamiltonian system described by equations
(\ref{hamqp}) with
$$
T_{ij} = m\,\delta_{ij},~~~~~~~~\theta_{ij}=\epsilon_{ij}\,eB,
~~~~~~~~V_{ij}=0.
$$

However, it is also well known \cite{peierls}, \cite{landau} that one can define a
set of noncanonical phase space variables which give rise the same
equations of motion.  In  concrete, consider  the Hamiltonian system
defined in terms of the Poisson Brackets relations of variables
$\{q,u\}$ as follows

\begin{eqnarray}
\label{landau2} [q^i,q^j]=0,~~[q^i,u_j]=\delta^i_j,&&[u_i,u_j]=-eB
\epsilon_{ij},\nonumber
\\
~~[u_i,u_3]=0,~~ H &=& \frac{m}{2}\left( u_1^2 ++u_2^2 +
u_3^2\right).\nonumber
\end{eqnarray}

Matrices $T, V,\theta$ are defined as before, so that the previous
Hamiltonian system is equivalent to (\ref{hamqu}).

Clearly, both  systems are  not connected by  canonical
transformations and might, in principle, give rise to inequivalent
quantum theories.

In what follows we discuss the noncommutative harmonic oscillator.

\subsection{Noncommutative harmonic oscillator}
In \cite{np}  the quantum  mechanics of the  harmonic oscillator  in
a fully noncommutative space, {\it i.e}, a space where coordinates
commutators and momenta commutators do not vanish, has been
discussed.

From the point of view of  the present article, the starting point
are the classical EoM of the system which we will write in terms of
variables $\{q,u\}$
\begin{eqnarray}
\label{ncnp}
\dot{q}^i &=&\delta^{ij}u_j, \nonumber
\\
\dot{u}_{i} &=& (B + \theta \omega^2){\epsilon_i}^j u_{j} -
(1-B\theta)\omega^2 \delta_{ij}q^j
\end{eqnarray}
with $\{i,j\} \in \{1,2\}$.

For this set of equations at least two Hamiltonian structures may be
defined. One of them is the following
\begin{eqnarray}
\mbox{[}q^i,q^j\mbox{]}
=0,~~~\mbox{[}q^i,u_j\mbox{]}=\delta^i_j,&&\mbox{[}u_i,u_j\mbox{]} =
(B+\theta\omega^2) \epsilon_{ij} \nonumber
\\
\label{h1}
H_1 = \frac{1}{2}(u_1^2+u_2^2) &+& \frac{\omega^2}{2}(1-B\theta)(q_1^2 +
q_2^2). \nonumber
\\
&&
\end{eqnarray}

There  is a  second order  Lagrangian from which  these  EoM  can be
derived as it can be readily seen by comparing  them with  the
Hamiltonian structure (\ref{hamqu}). Note that  $T_{ij}\neq V_{ij}$.
Besides, det$(\openJ_1) =1$ and therefore the Poisson Brackets
matrix is nowhere singular.\\

Another Hamiltonian structure for the same set of  EoM is the
following (we preserve previous notation for comparison purposes)
\begin{eqnarray}
\label{h2}
\mbox{[}q^i,q^j\mbox{]}& =&\theta \epsilon_{ij},
~~~~~~~~~~~\mbox{[}q^i,u_j\mbox{]}=
(1+\theta^2\omega^2)\,\delta^i_j,\nonumber
\\
&&\mbox{[}u_i,u_j\mbox{]} =
(\theta^3\omega^4+2\theta\omega^2+B)\,\epsilon_{ij} \nonumber
\\
H_2 &=& \frac{1}{2}(u_1^2+u_2^2)+\frac{\omega^2}{2}(1 + \theta^2 \omega^2)
((q^1)^2 + (q^2)^2)\nonumber
\\
&+& \omega^2 \theta (q^1u_2 - q^2 u_1).
\end{eqnarray}

In  this case, one  can check  that the  Poisson Bracket  matrix has
a singularity in parameter   space  for   $\theta  B =1$ since
det$(\openJ^{\nc})=(1-\theta B)^2$.

For completeness, let us write the classical Hamiltonian structure
which leads,  upon quantization, to the noncommutative harmonic
oscillator (using the notation of previous sections)
\begin{eqnarray}
\mbox{[}q^i,q^j\mbox{]}&            =&\theta            \epsilon_{ij},
~~~~\mbox{[}q^i,v_j\mbox{]} = \delta^i_j,\nonumber~~~
\mbox{[}v_i,v_j\mbox{]} =B\,\epsilon_{ij} \nonumber
\\
H^\nc&=&\frac{1}{2}\left((v_1)^2 + (v^2)^2 +
\omega^2((q^1)^2+(q^2)^2)\right)
\end{eqnarray}

Clearly,  this Hamiltonian  structure has  $T_{ij} =  V_{ij}$, up to
time rescaling. There is  no second order Lagrangian for this system
because coordinates have non vanishing Poisson Brackets relations
\cite{sh}, \cite{hs}. Moreover, the Poisson Bracket structure is
singular for $\theta\,B=1$.

In  summary, the  classical equations  of  motion under study can be
derived from a Hamiltonian system which may be obtained from a
second order Lagrangian or, from a Hamiltonian system which is not
derivable from a second order Lagrangian because coordinates have
non vanishing Poisson Brackets relations. In this last case, space
turns out to be noncommutative after quantization.

In the next section we study the quantization of these systems.

\section{Quantum Mechanics}
In  this section  we will  calculate explicitly  the energy levels
of the systems previously discussed.  For the case of Landau Levels,
as well as the noncommutative harmonic oscillator, these results are
very well known and we  will limit ourselves just to show the
results in order to compare with the ones obtained for the new
cases.

\subsection{Landau Levels}
Quantization of  (\ref{hlandau}), once  restricted to the  plane
$p_3$ constant, gives rise to an energy spectrum known as Landau
Levels. The energy levels of this Hamiltonian system are \cite{landau}
\begin{equation}
\label{ll} E^{\Landau}_\ell = \frac{{p_3}^2}{2m}
+\omega_0\,(\ell+1/2),
\end{equation}
with  $\omega_0  = eB/m$, $\hbar=1$, $c=1$ and  $\ell  =
0,1,2\cdots$.

The wave function is
\begin{equation}
\label{llw} \psi_\ell(q)       \propto
e^{-\frac{m\omega}{2}(q^2-q^2_0)q^2 }H_\ell\left(\sqrt{m\omega
}(q^2-q^2_0)\right),
\end{equation}
where $q^2$ is one of the coordinates on the plane spanned by
coordinates ($q^1,\ q^2$), $q^2_0$ is a constant  and $H_\ell(q)$ is
the Hermite polynomial of order $\ell$.

The quantum mechanics of Hamiltonian system (\ref{landau2}) gives
the same results as the  usual case previously summarized. Indeed,
it is enough to note that a realization of the commutators algebra
is given by the usual coordinate basis $q^1,q^2$ and the following
operators $u_i$
\begin{equation}
\label{urealization} u_i = -i \partial_{i} -e A_i,
\end{equation}
which,   once  implemented  into   the  Hamiltonian,   reproduces  the
Hamiltonian in variables $\{q,p\}$.

Therefore, the two Hamiltonian systems, upon  quantization give rise
to the same quantum theory. This is a non trivial result because,
even if  both  systems  are  related   by  a  rather trivial
relation  such as (\ref{urealization}), the transformation
$\{q^i,p_j\} \leftrightarrow \{q^i,u_j\}$ is not canonical.

\subsection{Noncommutative harmonic oscillator}
Let  us  consider the  Hamiltonian  system  (\ref{h1}).  The algebra
of commutators turns out to be
$$
\mbox{[}q^i,q^j\mbox{]}    =0,~~~\mbox{[}q^i,u_j\mbox{]}=i\delta^i_j,~~
\mbox{[}u_i,u_j\mbox{]} =i (B+\theta\omega^2) \epsilon_{ij} \nonumber
$$
which has a realization in coordinate representation $\{q^1,q^2\}$
$$
u_i = -i \nabla_{q^i} + \frac{1}{2}(B+\theta\omega^2)\epsilon_{ij}q^j
$$
By doing that, the Schr\"odinger equation of the system is
\begin{eqnarray}
 \left[-\frac{1}{2}\nabla^2_q + \frac{1}{2}\Omega^2{\bf q}^2
   + \frac{i}{2}\lambda(q^1\partial_2-q^2\partial_1)\right]\psi(q)
 &=& E\psi(q) \nonumber
\\
&&
\end{eqnarray}

with
$$
\Omega^2= \omega^2 + \frac{1}{4}(B-\theta\omega^2)^2,~~~~~
\lambda= (B+\theta\omega^2).
$$

 This equation can be solved completely.  In
order to do that it is convenient to parameterize the coordinate
space $\{q^1,q^2\}$ in polar coordinates $(r,\varphi)$. The
normalizable wave function turns out to be
\begin{equation}
\label{wave1}
\psi^\ell_n(r,\varphi)=e^{-\frac{\Omega^2}{2}r^2}e^{i\ell\varphi}\,L^{\ell}_n(\Omega r^2).
\end{equation}
with      $L^\ell_n$,     the      Legendre's      Polynomials     and
$\ell,n=0,1,\cdots$. The Hamiltonian eigenvalues are
\begin{equation}
E_{n,\ell} = \Omega \left(n+\ell + 1 \right) - \frac{\ell\lambda}{2}.
\end{equation}

This is  different from the noncommutative  harmonic oscillator
spectrum, which is given in terms of the  spectra of two harmonic
oscillators with different frequencies.

Finally, let us discuss the  quantum mechanics of $H_2$.  Consider the
canonical  coordinate  representation.  Let  $x_1,x_2$  be  the  space
coordinates  and $p_j$  the canonical  momenta, which  in coordinate
representation          has         the          standard         form
$p_j=-i\partial_{x_j}\equiv-i\partial_j$.  Then, a realization  of the
quantum commutators version of
(\ref{h2}) is
\begin{eqnarray}
q^1 &=&\sqrt{\theta}\,x_1, ~~~~~~ q^2 =\sqrt{\theta}\, p_1,
\\
u_1 &=&\sqrt{\frac{\theta B - 1}{\theta}}\,x_2 +
\frac{1+ \theta^2\omega^2}{\sqrt{\theta}}\, p_1
\\
u_2 & = &\sqrt{\frac{\theta B -1}{\theta}}\, p_2-
\frac{1+\theta^2\omega^2}{\sqrt{\theta}}\,x_1.
\end{eqnarray}
for $\theta > 0$ and $B\theta >1$.

The Hamiltonian turns out to be
\begin{eqnarray}
\label{hamcan2}
2\theta\, H_2 &=& (1+\theta^2\omega^2)(p_1^2 +x_1^2)
+ (\theta B-1)(p_2^2 + x_2^2) - \nonumber
\\
&-&4\sqrt{\theta B -1}\,L_3
\end{eqnarray}
with $L_3= x_1p_2-x_2p_1$.

This Hamiltonian corresponds to an anisotropic harmonic oscillator
with an angular momentum term. Indeed, it has the following
structure
$$
H = U(p_1^2 + x_1^2) +V (p_1^2 + x_1^2) +W\,L_3
$$
for $U,V,W$,  constants, and it  is always possible  to put it  in the
form
$$
H  =  \frac{1}{2M}  {\bf  p}^2  +  \frac{1}{2}M  (\Omega_1^2  x_1^2  +
\Omega_2^2x_2^2) + W\,L_3
$$
This is  not a diagonalizable system  as it was  shown in
\cite{we3}, and therefore  the quantum mechanics  it describes, is
different from the two cases before analyzed.\\

Furthermore, in this case the wave function cannot, even in
principle, be written in terms of the coordinates (because they do
not commute). There is no way to compare the results obtained for
this quantum system with the one described by commuting coordinates.

\section{Conclusions and Outlook}

In this paper we deal with one classical set of second order
equations of motion and we approach the construction of Hamiltonian
structures in three different ways.

We start from a second order Lagrangian for the system under
consideration and construct its Hamiltonian structure in the usual
way using two different sets of phase space variables $\{q, p \}$
and $\{q, u (\equiv \dot q) \}$. The Poisson brackets relations and
the Hamiltonian functions for both sets of phase space coordinates
are exhibited. The quantum theories are worked out and they turn out
to be equivalent.

Nevertheless, when the oscillator is described in terms of non
commuting coordinates, the quantum theory has a spectrum which is
different from the one previously found. Furthermore, its wave
functions cannot even be compared to the ones obtained when using
commuting coordinates. In this context, see also \cite{sh6}.

\bigskip
\noindent\textbf{Acknowledgements}: This  work was supported by grants from
FONDECYT-Chile grant-1095106, 1100777.

\end{document}